\documentclass[fleqn,twoside]{article}
\newcommand{\golempreprintnumbers}{%
FREIBURG-PHENO-2010-021, %
IPPP/10/51, %
DCPT/10/102}

\usepackage{espcrc2}
\usepackage{graphicx}
\usepackage{amsmath}

\newcommand{\ket}[1]{\left\vert{#1}\right\rangle}

\newcommand{\GOLEM}{Golem}
\newcommand{\GOLEMVC}{\texttt{golem95}}
\newcommand{\GOLEMXX}{\texttt{golem-2{.}0}}
\newcommand{\HAGGIES}{\texttt{haggies}}

\newcommand{\FORTRANVC}{\texttt{Fortran\,95}}
\newcommand{\FORM}{\texttt{Form}}
\newcommand{\SAMURAI}{\texttt{Samurai}}
\newcommand{\WHIZARD}{\texttt{WHIZARD}}

\title{
\vspace{-3em}{\small\hbox{}\hfill\golempreprintnumbers}\\[2em]
Recent Progress in the \GOLEM{} Project}

\author{Gavin Cullen\address[UoE]{%
	School of Physics and Astronomy, %
	The University of Edinburgh,
	Edinburgh EH9\,3JZ, UK}, %
	Nicolas Greiner\address[UoI]{%
	Department of Physics, %
	University of Illinois at Urbana-Champaign, %
	Urbana IL, 61801, USA},
	Alberto Guffanti\address[UFBr]{%
	Physikalisches Institut, %
	Albert-Ludwigs-Universit\"at, %
	79104~Freiburg, Germany}, %
	Jean-Philippe Guillet\address[LAPTH]{%
	LAPTH, %
	74941 Annecy le Vieux Cedex, France}, %
	Gudrun Heinrich\address[IPPP]{%
	IPPP, %
	University of Durham, %
	Durham DH1\,3LE, UK}, %
	Stefan Karg\address[RWTH]{%
	Institut f\"ur Theoretische Teilchenphysik und Kosmologie, %
	RWTH Aachen University, %
	52056~Aachen, Germany}, %
	Nikolas Kauer\address[RHUL]{%
	Department of Physics, %
	Royal Holloway, University of London, %
	Egham TW20\,0EX, UK}, %
	Tobias Kleinschmidt\addressmark[IPPP], %
	Eric Pilon\addressmark[LAPTH], %
	{\bf Thomas Reiter}\address[Nikhef]{
	Nikhef, %
	1098~XG Amsterdam, The Netherlands}, %
	J\"urgen Reuter\addressmark[UFBr], %
	Mark Rodgers\addressmark[IPPP], %
	Ioan Wigmore\addressmark[UoE]
}       
\begin{document}


\begin{abstract}
We report on the current status of the \GOLEM{} project which aims
at the construction of a general one-loop evaluator for matrix elements.
We construct the one-loop matrix elements from Feynman diagrams in a
highly automated way and provide a library for the reduction and
numerically stable evaluation of the tensor integrals involved in this
approach. Furthermore, we present applications to physics processes relevant
for the~LHC.
\vspace{1pc}
\end{abstract}


\maketitle

\section{Introduction}
\label{sec:golem:Introduction}

Precise predictions for signals and backgrounds at high energy colliders
are required to be able to identify and study New Physics in current experiments.
Processes with multi-particle final states will play a significant role in this task, 
in particular at the LHC, which emphasises the need for predictions at 
next-to-leading order~(NLO) for such~processes.

One of the challenges of NLO calculations is the numerically stable evaluation and
integration of the virtual corrections. Both the development of new
unitarity based methods and the improvement of traditional methods
based on Feynman diagrams have led to important new NLO results, 
in particular to physical predictions for $2\to 4$ processes at hadron 
colliders~\cite{Bredenstein:2008zb,Berger:2009zg,Bredenstein:2009aj,KeithEllis:2009bu,Berger:2009ep,Bevilacqua:2009zn,Binoth:2009rv,Bevilacqua:2010ve,Berger:2010vm,Bredenstein:2010rs,Ellis:2009zw}.
A recent overview can also be found in~\cite{Binoth:2010ra,Bern:2008ef}.

As these calculations are technically involved, it is important to 
spend some effort on two features: {\it automatisation} and 
{\it modularity}.
The latter allows for a combination of
automated  programs to calculate the virtual corrections 
with existing  tools for tree level matrix element generation,  
phase space integration and automated tools for 
the subtraction of soft/collinear singularities from 
real radiation via a standard interface~\cite{Binoth:2010xt}.

The \GOLEM{} approach to the calculation of one-loop 
amplitudes puts particular emphasis on automatisation, as the setup 
can be used for massless as well as massive particles and all types of spin 
structures. The Feynman diagrammatic approach implies that 
the rational parts of the amplitudes can be evaluated at no extra cost.
The reduction of the tensor integrals is done semi-numerically, thus avoiding 
large algebraic expressions.
A special choice of the integral basis preserves numerical stability 
in phase space regions with small Gram determinants. 
More details about the method are given in sections \ref{sec:golem:Algorithm}
and~\ref{sec:golem:Golem95}.

The \GOLEM{} method in combination with 
automated tools for the real radiation part 
has been used to obtain results for e.g. $q\bar{q}\rightarrow b\bar{b}b\bar{b}$, 
$pp\rightarrow ZZ+\mathrm{jet}$ and $pp\rightarrow \mathrm{graviton}+\mathrm{jet}$, 
which will be described in sections \ref{sec:golem:4b} and \ref{sec:golem:VVj}.

\section{Overview of the Algorithm}
\label{sec:golem:Algorithm}
For the consistent description of a cross-section for a
$2\rightarrow N$ particle process at NLO 
one needs to include the square of the tree level matrix element
$A^{(0)\dagger}_{N}A^{(0)}_{N}$, the real emission of an additional
massless particle $A^{(0)\dagger}_{N+1}A^{(0)}_{N+1}$ and the virtual
corrections consisting of the interference term of one-loop and tree-level
diagrams of the $2\rightarrow N$ process,
$A^{(0)\dagger}_{N}A^{(1)}_{N}+h.c.$

The \GOLEM{} project focuses on the efficient and numerically stable
computation of the virtual corrections, as the other contributions can
be added by interfacing with existing tools through a well defined generic
interface~\cite{Binoth:2010xt}.
The one-loop amplitude $A^{(1)}_N$ is represented as the
helicity projections of a sum of Feynman diagrams. Each diagram
$\mathcal{G}_a$ contributes
to one or more colour structures and can be written as a sum over tensor
integrals:
\begin{multline}
\mathcal{G}_a=\sum_{c\in\text{colour}}\ket{c}
\sum_{p=0}^r T_{c;p}^{\mu_1\ldots\mu_p}\times\\
\int\!\!\frac{\mathrm{d}^nq}{(2\pi)^n}
\frac{q_{\mu_1}\cdots q_{\mu_p}}%
{\prod_j ((q+r_j)^2-m_j^2+i\delta)}
\end{multline}
In our standard approach we use a Lorentz invariant decomposition of
the tensor integrals to separate the tensor structure from the associated
integral form factors. The fully contracted tensor structure is expressed
in terms of spinor products and the constants of the physics model, whereas
the Feynman parameter integrals are expressed in terms of
form factors
which can either be reduced further or be evaluated numerically.
An implementation of these form factors is available in the
library \GOLEMVC.~\cite{Binoth:2005ff,Binoth:2008uq}

As an alternative approach one provides the numerator function
\begin{equation}
\mathcal{N}_c(\hat{q},\mu^2) = \sum_{p=0}^r T_{c;p}^{\nu_1\ldots\nu_p}(\mu^2)
\hat{q}_{\nu_1}\cdots \hat{q}_{\nu_p},
\end{equation}
where $q^2=\hat{q}^2-\mu^2$, as an input to some reduction at the
integrand level~\cite{Ossola:2006us,Ossola:2007bb,Ossola:2007ax,%
Mastrolia:2008jb,Mastrolia:2010nb},
where $\mathcal{N}_c$ is evaluated for different
numerical values of $\mu^2$ and the four dimensional, complex vector $\hat{q}$.

\section{Reduction of the Tensor Integrals}
\label{sec:golem:Golem95}

The reduction of one-loop integrals follows an algorithm which has been developed in~\cite{Binoth:2005ff,Binoth:1999sp}, valid for an arbitrary number of
legs and both massive and massless particles. 
The ultraviolet and infrared divergences are regulated by dimensional regularisation. 

This algorithm has been implemented
as a \FORTRANVC{} library, \GOLEMVC,
for integrals with up to six external momenta~\cite{Binoth:2008uq}.
The library is publicly available at 
\texttt{lappweb.in2p3.fr/lapth/Golem/golem95.html}, 
together with detailed documentation and example programs.

The algebraic reduction of higher rank four-point  and
three-point functions to expressions containing only scalar  integrals 
necessarily leads to inverse 
Gram determinants appearing in the coefficients of
those scalar integrals. These determinants can become arbitrarily small 
in some phase space regions and
can therefore hamper a numerically stable evaluation of the one-loop
amplitude upon phase space integration. 
In order to deal with such situations in an automated way, 
our algorithm is able to avoid the occurrence of 
inverse Gram determinants completely. 
This is done by testing the value 
of the determinant and, in the case when it is smaller than 
a certain value, stopping the reduction {\it before} 
such inverse determinants are generated. 
In these cases, the tensor integrals, corresponding to integrals with
Feynman parameters in the numerator,  are evaluated by means of numerical
integration. The use of one-dimensional integral representations
guarantees a fast and stable~evaluation.
For integrals with internal masses, the option to evaluate the tensor integrals 
numerically prior to reduction, 
in regions where the Gram determinant tends to zero, is not yet 
supported, but is under construction.
For the IR finite scalar
box and triangle functions with internal masses, we use the expressions 
from the program {\tt OneLOop}~\cite{vanHameren:2009dr}.
The inclusion of an option to allow complex masses is underway.

To summarise, the library 
contains the following features:
\begin{itemize}
\item all tensor coefficients up to rank six six-point integrals, for massless as well as massive integrals
\item all scalar boxes, triangles, bubbles, tadpoles
\item boxes in $n+2$ dimensions up to rank three ($n=4-2\epsilon$) and in $n+4$
 dimensions up to rank one, and triangles in $n+2$ dimensions up to rank one (larger ranks of the higher dimensional integrals are not needed in the reduction).
\item automated detection of small Gram determinants and fast (one-dimensional)
numerical integration of tensor integrals to avoid the  occurrence of small inverse 
determinants
\item option to calculate the rational parts only.
\end{itemize}

\section{Automated Construction of One-Loop~Amplitudes}
\label{sec:golem:Golem20}
For the automated construction of the matrix elements we have developed a
package (\GOLEMXX~\cite{Binoth:2010pb})
combining \texttt{QGRAF}~\cite{Nogueira:1991ex} as a
diagram generator with the symbolic manipulation program
\FORM~\cite{Vermaseren:2000nd} for the algebraic simplification
of the generated amplitude. The expressions for the helicity projections
of each diagram contributing to the process are further processed by
\HAGGIES~\cite{Reiter:2009ts} to produce optimised \FORTRANVC{} code
for a fast numerical evaluation. A large part of the infrastructure and
the user interface is implemented as a \texttt{Python} program.

For the reduction of the tensor integrals \GOLEMXX{} supports two choices.
One can express the amplitude in terms of integral form factors as defined
in~\cite{Binoth:2005ff} and implemented in \GOLEMVC{}~\cite{Binoth:2008uq}.
Alternatively, the program can generate the numerator of each Feynman diagram
in a form suitable as input for \SAMURAI{}~\cite{Mastrolia:2010nb}.
For each of the two options the amount of human intervention is limited
to setting up an appropriate configuration file.

In order to further improve the applicability of the program we are currently
implementing an interface to FeynRules~\cite{Christensen:2008py} model files.
Interoperability with existing Monte Carlo programs will be achieved by
providing an interface conforming to the Binoth-Les Houches
Accord~\cite{Binoth:2010xt}
and some additional functionality to provide the relevant information about
the matrix element to the \texttt{POWHEG BOX}~\cite{Alioli:2010xd}.

The described features offer a great flexibility to the potential user;
the achieved performance of the matrix element code and its modularity
allow for a wide range of applications.
\section{Results for
$q\bar{q}\rightarrow b\bar{b}b\bar{b}$ 
}
\label{sec:golem:4b}
We have used the setup described in Section~\ref{sec:golem:Golem20} for
generating a \FORTRANVC{} implementation of the one-loop matrix element
for the process $q\bar{q}\rightarrow b\bar{b}b\bar{b}$~\cite{Binoth:2009rv},
which is one of the channels of the full $b\bar{b}b\bar{b}$ cross-section
at the LHC.

The QCD induced contribution to the $4b$-signature constitutes an important
background for Higgs searches in the MSSM, where for certain parameter
regions the heavy Higgs boson decays predominantly into
$b\bar{b}$~pairs~\cite{Bern:2008ef,Lafaye:2000ec}.
Also outside the framework of the MSSM interesting physics scenarios
such as certain \textit{hidden valley} models lead to collider signatures
with many $b$-quarks~\cite{Krolikowski:2008qa}.

We have ensured the correctness of our one-loop matrix element by comparison
to a second, independent setup using FeynArts~\cite{Hahn:2000kx},
\FORM~\cite{Vermaseren:2000nd} and Maple as a tool chain.
Integrated results have been obtained by three different setups:
\begin{itemize}
\item the virtual matrix element obtained from \GOLEMXX{} has been used
   as a stand-alone program to reweight unweighted leading order events
   generated by \WHIZARD~\cite{Kilian:2007gr,Moretti:2001zz}. All tree
   like parts and the infrared subtraction terms
   as defined in~\cite{Catani:1996vz}
   using a slicing parameter~\cite{Nagy:2005gn} have been implemented
   in~\WHIZARD.
\item the virtual matrix element from \GOLEMXX{} has been linked
   into \texttt{MadEvent}~\cite{Maltoni:2002qb} together with the tree-like
   parts from \texttt{MadGraph}~\cite{Stelzer:1994ta}
   and the dipoles from \texttt{MadDipole}~\cite{Frederix:2008hu,Frederix:2010cj}.
\item the third setup is the same as the previous but using
   \SAMURAI~\cite{Mastrolia:2010nb} in place of \GOLEMVC~\cite{Binoth:2008uq}
   for the reduction of the numerators of the virtual matrix element.
\end{itemize}

The results use LHC kinematics ($\sqrt{s}=14\,\text{TeV}$ centre
of mass energy). As we calculate observables for well separated jets,
we cluster the final state particles by applying the
$k_T$-algorithm~\cite{Blazey:2000qt} into $b$-jets,
which then have to pass the
set of cuts,
\begin{equation}
\begin{array}{rcl}
p_T(b_j)&>&30\,\text{GeV},\\
\vert\eta(b_j)\vert&<&2{.}5,\\
\Delta R(b_i,b_j)&>&0{.}8.
\end{array}
\end{equation}
We use the parton density functions CTEQ6{.}5~\cite{Tung:2006tb}
with two-loop running of $\alpha_s$. The central renormalisation
scale is $\mu_0=\frac12\sqrt{\sum_{j=1}^4 [p_T(b_j)]^2}$ while the
factorisation scale has been fixed at $\mu_F=100\,\text{GeV}$.
The initial state includes $q\in\{u,d,s,c\}$; for the whole calculation
we use the approximation $m_b=0$ and~$m_t\rightarrow\infty$.

During phase space integration numerical stability was monitored
by checking the magnitude of the $K$-factor for a given phase space point
and also the quality of the cancellation of the infrared poles. In the worst
case about $0.5\%$ of the events needed evaluation in quadruple precision.
By using the setup with \SAMURAI, on top of previous tests
we could also use several reconstruction tests as described
in~\cite{Mastrolia:2010nb} for estimating the numerical quality for each
phase space~point. 

\begin{figure}[ht]
\includegraphics[width=\columnwidth]{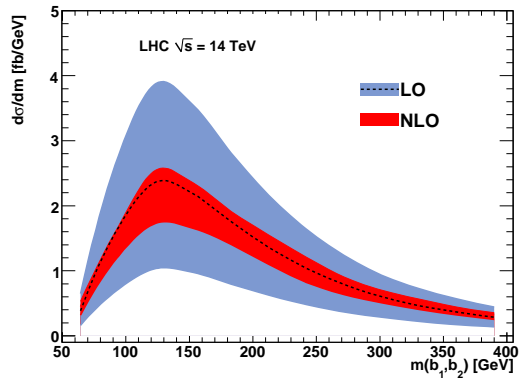}
\caption{Distribution of the invariant mass of the hardest $b$-jet pair
in $q\bar{q}\rightarrow b\bar{b}b\bar{b}$. The error bands represent a
variation of the renormalisation scale $\mu_R=x\mu_0$ for $x\in[\frac12;4]$.}
\label{fig:golem:4b-mb1b2}
\end{figure}

Figure~\ref{fig:golem:4b-mb1b2} shows the distribution for the invariant
mass of the hardest pair of $b$-jets, where the jets are ordered by
their~$p_T$. The error bands are obtained by a variation of the
renormalisation scale $\mu_R=x\mu_0$ for $x\in[\frac12;4]$.
The dashed line corresponds to the leading order result at $\mu_R=\mu_0$. 
As expected, the plot shows a clear reduction of the scale dependence.

\begin{figure}[ht]
\includegraphics[width=\columnwidth]{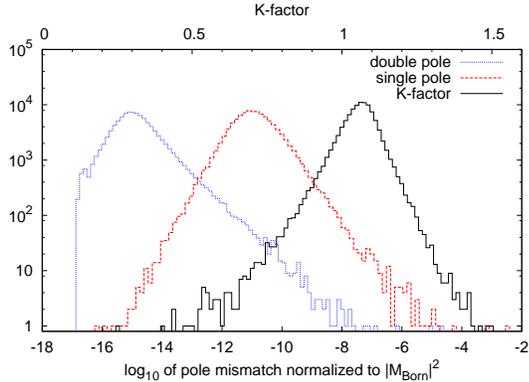}
\caption{Criteria used for the determination of numerical stability
of the virtual matrix element:
cancellation of the infrared poles (lower $x$-axis) and pointwise
$K$-factor (upper $x$-axis) for $10^5$ events.
The virtual part has been evaluated using \SAMURAI{} in double~precision.
}
\label{fig:golem:4b-samurai}
\end{figure}

In Figure~\ref{fig:golem:4b-samurai} we examine the precision of $10^5$
phase space points by three different criteria. Since the cancellation
of the single and double poles on their own are not a reliable indicator
for numerical problems we also consider the local $K$-factor as a heuristic
criterion (see~\cite{Reiter:2009kb}). In the chosen setup with \texttt{MadEvent} as
an integrator and \SAMURAI{} for the reduction of the amplitude
the number of unstable points in double precision is below one per mil.

\section{Results for
$pp\rightarrow VV+\mathrm{jet}$ 
and $pp\rightarrow G+\mathrm{jet}$ 
}
\label{sec:golem:VVj}

\begin{figure}[ht]
{\flushleft \includegraphics[width=7.5cm,angle=0,clip=true]{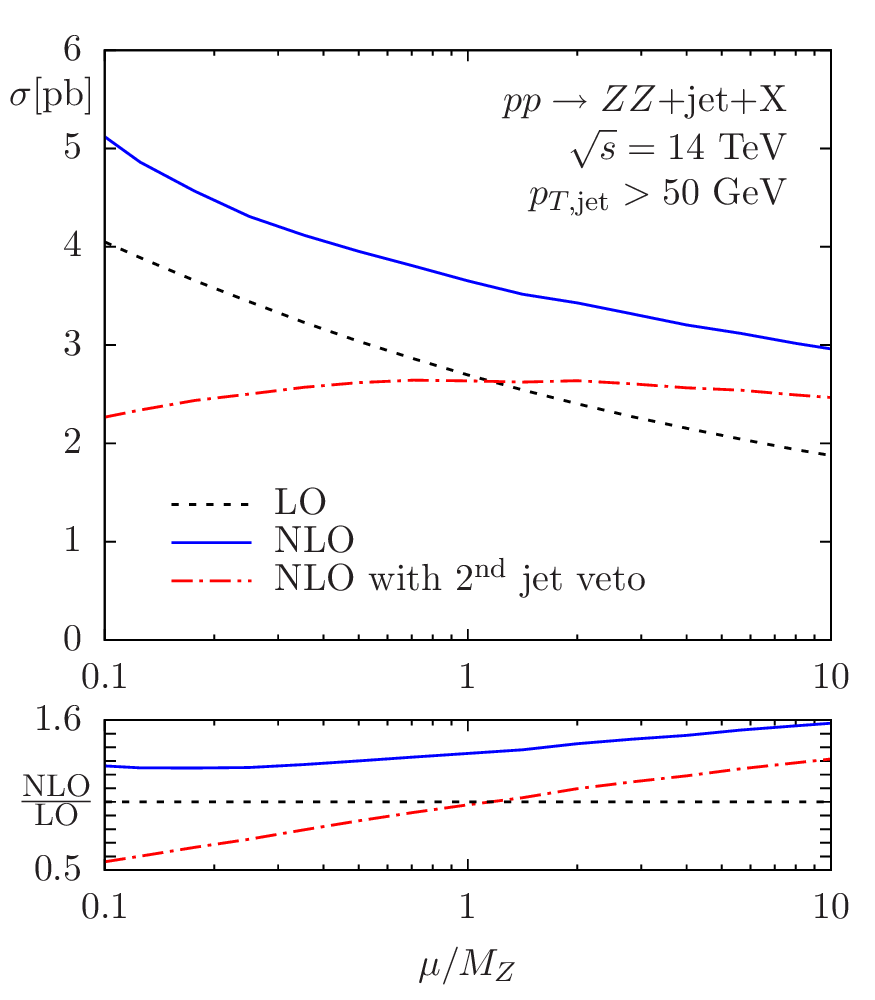}}
\caption{%
Scale dependence ($\mu_R=\mu_F=\mu$) of the $ZZ$+jet cross section at the LHC 
($\sqrt{s}=14$ TeV) with $p_{T,\,\textrm{jet}} > 50$ GeV for the hardest jet 
in LO (dotted) and NLO (solid).  The exclusive NLO cross section when a
$p_{T,\textrm{jet}} > 50$ GeV veto for additional jets is applied is also
shown (dot-dashed).
\label{fig:golem:zzj}}
\end{figure}

\begin{figure}[ht]
{\flushleft \includegraphics[width=7.5cm,angle=0,clip=true]{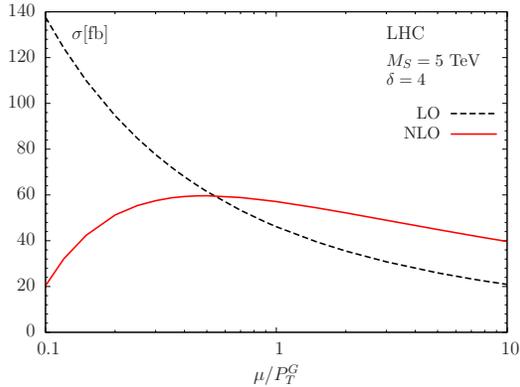}}
\caption{%
Scale dependence ($\mu_R=\mu_F=\mu$) of the graviton+jet cross section at the LHC 
($\sqrt{s}=14$ TeV) in LO (dotted) and NLO (solid) in a Kaluza-Klein model with 
$\delta=4$ extra dimensions and a fundamental scale $M_S=5$ TeV.  Selection cuts are 
described in the main text.
\label{fig:golem:gj}}
\end{figure}

{\GOLEM} tensor reduction methods have also been used to compute
fully-differential 
NLO QCD corrections to $ZZ$+jet \cite{Binoth:2009wk} and Kaluza-Klein 
graviton+jet 
\cite{Karg:2009xk} production at the LHC and Tevatron.  At these colliders, 
graviton+jet production is an important channel for graviton searches and $ZZ$+jet 
production is an important background for Higgs particle and new physics searches.
In both calculations, we have identified the renormalisation and factorisation scales:
$\mu_R=\mu_F=\mu$.  In Figs.~\ref{fig:golem:zzj} and \ref{fig:golem:gj} we show the scale
dependence of LHC cross sections for $ZZ$+jet and graviton+jet production,
respectively.  The central choice for $\mu$ is the $Z$ mass and the transverse 
momentum of the graviton $p^G_T$, respectively.  The graviton search selection 
requires $p_T^\text{miss} > 500$ GeV and jets are required to satisfy $p^j_T > 50$ GeV
and $|\eta_j|<4.5$.
For $ZZ$+jet production, we require $p_T > 50$ GeV for the hardest jet.  Following 
Ref.~\cite{Dittmaier:2007th}, in Fig.~\ref{fig:golem:zzj} we also show the NLO cross section 
when 2-jet events with a second hardest jet with $p_T > 50$ GeV are vetoed.  The jet 
veto clearly reduces the residual scale variation.
We employed the CS dipole subtraction \cite{Catani:1996vz} implementations in
\texttt{MadDipole} \cite{Frederix:2008hu} and \texttt{SHERPA} \cite{ZZj_Sherpa}
to calculate numerical results for the finite real corrections contribution. 
For both calculations internal cross checks and comparisons have been carried out (see 
\cite{Binoth:2009wk,Karg:2009xk}).  Our $ZZ$+jet calculation has also been validated 
through a tuned comparison with the independent calculation of Ref.~\cite{DKU_ZZj}. 
For a fixed phase-space point, the virtual corrections, obtained using different 
calculational techniques,  agree at the level of $10^{-8}$ or better. The comparison 
of full NLO cross sections for the LHC and Tevatron, shows agreement at the per mil 
level.  Details are reported in Ref.~\cite{Binoth:2010ra}.

\section{Conclusion}
\label{sec:golem:Conclusion}
We have given an overview of the recent developments in the \GOLEM{}
collaboration whose main goals are the automatisation of the computation
of one-loop matrix elements and the provision of precise predictions
for LHC observables.
As an integral part of this work we developed a library
(\GOLEMVC) of form factors for the reduction of one-loop tensor integrals,
which in its latest version contains all integrals for
one to six propagators up to full rank both for massless and massive
particles in dimensional regularisation. We have implemented the
\GOLEM{} method for computing one-loop amplitudes in a highly automated
framework (\GOLEMXX) which has been applied to the computation of the
quark-induced channel of the process $q\bar{q}\rightarrow b\bar{b}b\bar{b}$.
A similar method using the \GOLEM{} tensor reduction has been used to
compute the production of $Z$-pairs associated with a jet and the production
of a graviton plus one jet at the LHC.

\section*{Acknowledgements}
N.G.~has been supported by the U.~S.~Department of Energy
   under contract No.~DE-FG02-91ER40677. 
G.H.,~T.K. and M.R. have been supported by the British
   Science and Technology Facilities Council~(STFC).
N.K.~gratefully acknowledges the support of the IPPP Durham through
   an~Associateship.
T.R.~has been supported by the Foundation FOM, project FORM~07PR2556.
J.R.~has been partially supported by the Ministery of Science and Culture
(MWK) of the German state Baden-W\"urttemberg.


\end{document}